\documentclass[a4paper,11pt]{article}
\usepackage{jcappub} 
\usepackage{comment}
\usepackage{physics}
\usepackage{todonotes}
\usepackage{dsfont}
\usepackage{slashed}
\usepackage{hyperref}
\usepackage{url}
\usepackage{subfig}

\title{
\huge{Hubble-Induced Phase Transitions: Gravitational-Wave Imprint of Ricci Reheating from Lattice Simulations}}
\author[a,b]{Dario Bettoni,}
\emailAdd{dbet@unileon.es}
\author[c]{Giorgio Laverda,}
\emailAdd{giorgio.laverda@tecnico.ulisboa.pt}
\author[d,e]{Asier Lopez-Eiguren, }
\emailAdd{asier.lopez@ehu.eus}
\author[f]{Javier Rubio}
\emailAdd{javier.rubio@ucm.es} 
\vspace{5cm}
\affiliation[a]{Departamento de Matemáticas, Universidad de León, Escuela de Ingenierías Industrial, Informática y Aeroespacial Campus de Vegazana, s/n 24071 León, Spain}

\affiliation[b]{Instituto Universitario de F\'isica Fundamental y Matem\'aticas~(IUFFyM), Universidad de Salamanca, Salamanca, E-37008, Spain}

\affiliation[c]{Centro de Astrof\'{\i}sica e Gravita\c c\~ao  - CENTRA,
Departamento de F\'{\i}sica, Instituto Superior T\'ecnico - IST,
Universidade de Lisboa - UL,
Av. Rovisco Pais 1, 1049-001 Lisboa, Portugal} 

\affiliation[d]{Department of Physics, UPV/EHU, 48080, Bilbao, Spain}
\affiliation[e]{EHU Quantum Center, University of Basque Country, UPV/EHU}

\affiliation[f]{Departamento de Física Teórica and Instituto de Física de Partículas y del Cosmos (IPARCOS-UCM), Universidad Complutense de Madrid, 28040 
Madrid, Spain} 

\abstract{Gravitational waves offer an unprecedented opportunity to look into the violent high-energy processes happening during the reheating phase of our Universe. We consider a Hubble-induced phase transition scenario as a source of a post-inflationary stochastic background of gravitational waves and analyse the main characteristics of its spectrum for the first time via  numerical methods. The output of a large number of fully-fledged classical lattice simulations is condensed in a set of parametric formulas that describe key features of the GW spectrum, such as its peak amplitude and characteristic frequency, and avoid the need for further time-consuming simulations. The signal from such stochastic background is compared to the prospective sensitivity of future gravitational-wave detectors.}
\keywords{physics of the early universe, gravitational waves, kination, symmetry breaking, lattice simulations}
\preprintnumber{}
\begin{document}
\maketitle

\section{Introduction} \label{sec:intro}

The early history of our Universe has a large gap. On one side, cosmological observations provide us with information about the initial quasi-de Sitter expansion through CMB measurements. On the other side, a combination of astrophysical and particle-physics measurements outlines the so-called hot Big Bang, that is to say the physical processes that lead to nucleosynthesis, and, ultimately, the large-scale Universe as we see it today. The period that goes from the end of inflation to the beginning of Big Bang nucleosynthesis is a vast unknown outside the reach of current telescopes and particle accelerators. The only messenger capable of delivering us information about this primordial dark age is gravitational waves. Since the revolutionary detection of black hole binary mergers in 2015 \cite{LIGOScientific:2016aoc, LIGOScientific:2016sjg, LIGOScientific:2017vwq} and especially after the detection of a stochastic background in 2023 \cite{NANOGrav:2023gor, NANOGrav:2023hvm, EPTA:2023fyk, EPTA:2023xxk, Ellis:2023oxs}, gravitational waves (GW) have become an incredible tool for studying the physics of the early Universe and the violent phenomena that characterise it. Therefore, it is crucial to thoroughly investigate the GW signal associated to phase transitions, particle production, topological defects and other non-homogeneous processes happening during the reheating phase, in order to evaluate a possible detection by present-day and future GW experiments.

In this work, we present an exhaustive description of the Stochastic Gravitational Waves Background (SGWB) produced by a Hubble-Induced Phase Transition (HIPT) \cite{Bettoni:2019dcw} taking place at the end of inflation. Such a phase transition is triggered whenever these two assumptions are realised:
\begin{enumerate}
    \item A phase of stiff expansion with $w>1/3$ follows the end of inflation;
    \item There exist a prototypical spectator field that is energetically subdominant and non-minimally coupled to the background curvature.
\end{enumerate}
In this setting, the time-dependence of the Ricci scalar induces the appearance of a large tachyonic mass for the spectator field at the beginning of the stiff expansion phase, acting like a ``cosmic clock" that sets off a crossover phase transition for the spectator field. Thanks to the ensuing tachyonic instability, the scalar field modes are exponentially amplified, thus leading to the production of large-amplitude GWs as the self-backreaction effects become strong and put an end to the tachyonic phase. Such scenario has been considered in several works as an efficient (re)heating mechanism in non-oscillatory models of inflation \cite{Bettoni:2021qfs, Bettoni:2021zhq, Laverda:2023uqv, Figueroa:2024asq, Opferkuch:2019zbd}, with applications to the Standard-Model Higgs \cite{Laverda:2024qjt} and the production of topological defects \cite{Bettoni:2018pbl}. It has also been considered for applications in Affleck-Dine baryogenesis \cite{Bettoni:2018utf}, in non-thermal first-order phase transition \cite{Kierkla:2023uzo, Mantziris:2024uzz} and for the late-time generation of neutrino masses \cite{Goertz:2024gzw}. However, a thorough description of the GW signal from such phase transition is still missing. The present work aims at filling this gap. Our approach is based on both analytical results and fully-fledged classical lattice simulations. In particular, we will compute the evolution equation for the metric tensor perturbations and the stress-energy tensor of a spectator field non-minimally coupled to curvature that sources it. The exact expression of the stress-energy tensor is instrumental in implementing the HIPT scenario in the publicly-available \texttt{$\mathcal{C}osmo\mathcal{L}attice$} code \cite{Figueroa:2020rrl, Figueroa:2021yhd}. The numerical output of 60 individual simulations is summarised in a set of easy-to-use parametric fitting formulas that eliminate the need for further time-consuming simulations. In particular, we focus on studying the typical amplitude and frequency at the peak of the spectrum, its integrated energy-density and its spectral shape. Notice that the resulting GW signal is inherently linked to the appearance of a large-amplitude anisotropic field distribution, a feature that cannot be described with the simplifying homogeneous field approximation assumed in earlier works \cite{Opferkuch:2019zbd, Dimopoulos:2018wfg, Figueroa:2016dsc, Fairbairn:2018bsw}. To the best of our knowledge, this is the first time that a parametric spectral shape is derived from lattice simulations for a crossover phase transition. Finally, we make a comparison between the signal and the sensitivity curves of several GW detectors and comment on the possibility of discovery.

The paper is structured as follows: in Section \ref{sec:overview} we introduce the HIPT scenario and present some of its core features. In Section \ref{sec:GW} we present the equation of motion of the transverse-traceless metric tensor perturbation and the stress-energy tensor sourcing GWs from the inhomogeneous spectator field.  In Section \ref{sec:lattice} we describe some details of the numerical lattice simulation and present some typical output. In Section \ref{sec:fitting_formulas} we compute the parametric formulas that summarise the characteristics of the GW signal. In Section \ref{sec:detectors} we compare the predicted signal with current proposals of GW detectors. Finally, in Section \ref{sec:conclusion} we make some remarks about the work, presenting also some possible future research directions.

\section{How to Hubble-induce a crossover phase transition}\label{sec:overview}

The HIPT scenario is based on a minimal number of assumptions that concern the field content and interactions in the fundamental Lagrangian. In particular, we consider two distinct sectors that are allowed to indirectly interact through gravity, namely an inflaton field $\phi$ and a prototypical scalar field $\chi$
\begin{equation}\label{eq:S}
\frac{\cal L}{\sqrt{-g}}=\frac{M_P^2}{2}R+{\cal L}_\phi+{\cal L}_\chi\,,
\end{equation}
with $M_P=2.44 \times 10^{18}$ GeV the reduced Planck mass and $R$ the Ricci scalar. The details of the inflationary stage are encoded in the Lagrangian density ${\cal L}_{\phi}$, which we will not specify in the following, as we only require that it produces a quasi de-Sitter phase ($w_{\phi}=-1$) followed by a \emph{kination} phase, i.e. a stiff expansion phase where the inflaton's equation of state parameter is $w_{\phi}=+1$. Such an expansion phase can be realised in a host of scenarios, ranging from asymmetric $\alpha$-attractors \cite{Akrami:2017cir, Dimopoulos:2017zvq,Dimopoulos:2017tud, Garcia-Garcia:2018hlc} to variable gravity settings \cite{Wetterich:1987fm,Wetterich:1994bg,Rubio:2017gty} or axion-like dynamics \cite{Gouttenoire:2021jhk}, but it is perhaps best described in terms of a non-oscillatory quintessential inflation model \cite{Peebles:1998qn,Spokoiny:1993kt} in which the kination phase interpolates between the early- and late-time accelerated expansion of the Universe (for a comprehensive review see 
\cite{Bettoni:2021zhq}). Regarding the spectator field Lagrangian density ${\cal L}_\chi$, we only assume it to be $\mathds{Z}_2$-symmetric and to contain a non-minimal interaction with curvature, namely
\begin{equation}\label{eq:lagchi}
\frac{{\cal L}_\chi}{\sqrt{-g}}=-\frac12 \partial^\mu\chi \partial_\mu \chi-\frac12 \xi R \chi^2-\frac{\lambda}{4} \chi^{4}\,. 
\end{equation}
The appearance of a non-minimal curvature coupling is a natural occurrence to be expected for quantum fields on curved backgrounds. Indeed, as the renormalisation procedure of the corresponding energy-momentum tensor involves the presence of such counter-terms, setting $\xi$ to vanish should be understood as an ad-hoc non-general choice viable only at a specific energy scale \cite{Birrell:1982ix, Mukhanov:2007zz}. To complete the fundamental picture, we assume the scalar field to be energetically subdominant (\emph{spectator} scalar field) with respect to the inflaton counterpart and in so doing, we neglect its backreaction effect on the background metric and the modification to the graviton propagator that would otherwise be present.

The Einstein equations following from the variation of the action \eqref{eq:S} with respect to the metric take the form 
\begin{equation}
G_{\mu\nu} = \frac{1}{M_{\rm P}^2}\left(T_{\mu\nu}^{\phi}+T_{\mu\nu}^{\chi}\right)\,,
\end{equation}
with $G_{\mu\nu}$ the Einstein tensor,  $T^\phi_{\mu\nu}$ the stress-energy tensor of the unspecified inflaton sector and
\begin{equation} \label{eq:energy_momentum_tensor}
T^\chi_{\mu\nu} = \partial_\mu\chi\partial_\nu\chi-g_{\mu\nu}\left(\frac12\partial_\alpha\chi\partial^\alpha\chi+\frac{\lambda}{4}\chi^4\right)+\xi\left(G_{\mu\nu}+g_{\mu\nu}\Box-\nabla_\mu\nabla_\nu\right)\chi^2\,,
\end{equation}
the one associated to the spectator field $\chi$.  Note that the contraction of $T^{\chi}_{\mu\nu}$ with arbitrary timelike vectors $u^\mu$ and $u^\nu$ is not necessarily positive definite at all times \cite{Ford:1987de,Ford:2000xg,Bekenstein:1975ww,Flanagan:1996gw}, opening the door to potential violations of the weak energy condition $T^{\chi}_{\mu\nu}u^\mu u^\nu \geq 0$ in this sector. Still, the overall energy density of the Universe is positive definite at all times.
The evolution of the spectator field is dictated by the Klein-Gordon equation
\begin{equation}
    \ddot{\chi} + 3H\dot{\chi} - \frac{1}{a^2} \nabla^2 \chi + \xi R(t) \chi + \lambda \chi^3 = 0 \, ,
    \label{eq:klein_gordon_spect}
\end{equation}
where dots denote derivatives with respect to the coordinate time $t$. The effective time-changing mass for $\chi$ depends on the evolution of the Ricci scalar. For a fixed flat Friedmann--Lema\^itre--Robertson--Walker metric (FLRW)  $g_{\mu\nu}={\rm diag}(-1,a^2(t)\,\delta_{ij})$ with scale factor $a(t)$, this quantity takes the form $R= 3(1-3w_\phi)H^2$ with $H = \dot{a}/a$ the Hubble rate and $w_\phi$ the effective inflaton equation-of-state parameter, identified with the overall background equation of state of the Universe during the early stages of our scenario. During inflation, the non-minimal coupling to curvature acts as a large mass term, safely confining the scalar field to its vacuum and preventing the formation of large isocurvature perturbations \cite{Bettoni:2019dcw, Laverda:2024qjt}. During kination, the Ricci scalar becomes negative since $w_\phi=1$ and triggers the spontaneous breaking of the $\mathds{Z}_2$ symmetry in the $\chi$ sector. Two degenerate true vacua appear then at large field values, but their position decreases monotonically towards zero as the Hubble function decreases with time. 

To study the dynamics of the spectator field in the kination phase, we assume the inflation-to-kination transition to be instantaneous as compared with the evolution of the scalar field \footnote{One could always embed the HIPT in a specific inflationary scenario with a model-dependent non-instantaneous transition to kination. This is particularly important whenever the tachyonic phase is not intense enough or when the non-minimal coupling parameter is small, namely $\nu<5$, so that a good separation of scales in the tachyonic band cannot be maintained \cite{Bettoni:2019dcw}. Such an approach was followed in \cite{Figueroa:2024asq} together with a choice of smaller quartic couplings $\lambda\in[10^{-15}, \; 10^{-4}]$ as compared to the typical Higgs-like values $\lambda\in[10^{-6}, \; 10^{-1}]$ in our study. Being $\nu\in[6, \; 24]$, our chosen parameter space realises strong phases of tachyonic particle production where an instantaneous transition to kination constitutes a very good approximation.} and parametrise the temporal dependence of the scale factor and the Hubble rate as 
\begin{equation}\label{eq:backkin}
a= a_{\rm kin}
\left[1+3 H_{\rm kin} \left(t-t_{\rm kin}\right)\right]^{1/3}\,, \hspace{10mm}  H=\frac{H_{\rm kin}}{1+3H_{\rm kin}(t-t_{\rm kin})}\,, 
\end{equation}
with $a_{\rm kin}$ and $H_{\rm kin}$ the values of these quantities at the onset of kinetic domination, arbitrarily defined at $t=t_{\rm kin}$. In terms of these quantities the energy density $\rho_\chi=T^\chi_{00}$ and pressure $p_\chi=\sum_k T^\chi_{kk}/(3a^2)$ of the spectator field following from \eqref{eq:energy_momentum_tensor} take the form 
\begin{eqnarray} \label{eq:energy_density_pressure}
\rho_{\chi} &=& \frac{\dot\chi^2}{2}+\frac{(\nabla\chi)^2}{2a^2}+3\xi\left( H^2+H\partial_t-\frac13\frac{\nabla^2}{a^2}\right)\chi^2 +\frac{\lambda}{4}\chi^4\,,\\
\nonumber
p_\chi & = &  \frac{\dot\chi^2}{2}-\frac{\vert\nabla\chi\vert^2}{6a^2}+3\xi\left(H^2-\frac13\left(\partial_t^2+2H\partial_t\right)+\frac29\frac{\nabla^2}{a^2}\right)\chi^2-\frac{\lambda}{4}\chi^4\,.
\end{eqnarray}
In the following, it will be convenient to consider dimensionless quantities, which can be achieved by normalising all dimensionful quantities, especially the scalar field $\chi$, by the kination scale $H_{\rm kin}$. We also introduce the conformal time variable $d\tau=dt/a$ which we then rescale to obtain
\begin{equation} 
    z = a_{\rm kin}\chi_{\rm kin}\tau \,, \hspace{5mm}
  \mathcal{H}(z)= \frac{a'}{a}=\frac{1}{2(z+\nu)} \,, \hspace{5mm}  a(z)=a_{\rm kin}\sqrt{1+\frac{z}{\nu}}\,,\hspace{5mm} \nu = \sqrt{\frac{3\xi}{2}} \,.
\end{equation}
with $\chi_{\rm kin}=\sqrt{6\xi}H_{\rm kin}$. Note that, instead of working with the coupling parameter $\xi$, we introduce a parameter $\nu$ following the conventions already used in \cite{Laverda:2023uqv, Bettoni:2019dcw, Bettoni:2021qfs, Bettoni:2021zhq}. The dynamics of the system can be split into two phases. The initial phase is defined by the non-minimal coupling term $\sim \nu^2 R \chi^2$ dominating the dynamics. Expanding the scalar field into quantum modes reveals that the amplitude of each mode undergoes an exponential tachyonic amplification described in terms of Bessel functions \cite{Bettoni:2019dcw}. These are the solutions to the mode equation in momentum space
\begin{equation}
    Y'' - \left( \textrm{k}^2 - M^2(z)\right)Y + \lambda Y^3 = 0\,.
    \label{eq:Yeom}
\end{equation}
with vacuum initial conditions. The time-dependent effective mass $M^2(z)=(4\nu^2-1){\cal H}^2$ defines the the band of tachyonic modes that grow exponentially at the beginning of kination. Notice that we have rescaled fields, momenta and spatial coordinated to their conformal version 
\begin{equation}
 Y = \frac{a}{a_{\rm kin}}\frac{\chi}{\chi_{\rm kin}} \,,  \hspace{10mm}
 \mathbf{y} = a_{\rm kin}\chi_{\rm kin}\mathbf{x} \, ,\hspace{10mm} 
    \textrm{k} = \frac{k}{a_{\rm kin}\chi_{\rm kin}} \,. 
    \label{eq:variablesredef_scalar}
\end{equation} 
In the linear phase, the original quantum nature of the system is quickly lost, thanks to the fast classicalisation of the tachyonically-produced particle excitations. This causes the appearance of amplified stochastic patches on typical scales $\sim \mathcal{H}^{-1}$  that grow until the self-interaction contribution $\sim \lambda \chi^4$ can no longer be neglected \cite{Felder:2001kt}. The time of \emph{backreaction} coincides with the completion of the first semi-oscillation, happening typically in less than one e-fold from the onset of kination. At this time, the field enters a non-linear phase characterised by the breaking of the locally-homogeneous patches. Contrary to the initial tachyonic phase, numerical methods are necessary to follow the self-interacting dynamics, since the contribution from spatial gradients and non-linear interactions cannot be longer ignored \cite{Laverda:2023uqv, Felder:2001kt}. Eventually, the system develops into a regime of turbulence \cite{Micha:2004bv} with a typical transfer of energy from IR modes to UV ones. In less than two e-folds, the average equation of state of the spectator field reaches $w_{\chi}=1/3$ and the following evolution of the scalar field's energy density can be tracked as that of a perfect relativistic fluid. The (re)heating of the Universe is ultimately achieved when the inflaton energy density decays below the scalar field's energy density. The timescale of heating phase is then completely defined in terms of the energy-density $\rho_\chi(z_{\rm rad})$ and the scale of kination $H_{\rm kin}$. 

Thanks to the parametric formulas obtained in \cite{Laverda:2023uqv}, we already possess a full characterisation of the heating phase derived from hundreds of lattice simulation, where the main quantities are defined as functions of $(H_{\rm kin},\, \nu, \,\lambda)$. In particular, the radiation time and the energy-density at radiation time are described by the simple formulas
\begin{equation}
    z_{\text{rad}}(\lambda, \nu) = \gamma_1 + \gamma_2 \, \nu \,,\hspace{5mm}
    \rho_{\text{rad}}(\lambda, \nu) = 16 H^4_{\rm kin} \, \exp\left({\delta_1 + \delta_2 \, \nu} +{\delta_3}\ln \nu\right) \,,
    \label{eq:fit_z_rad}
\end{equation}
with coefficients
\begin{align}
    &\gamma_1(\lambda) = 33.63 + 15.02 \, n - 0.22 \, n^2 \,,\hspace{2mm}
    \gamma_2(\lambda) = 7.91 - 0.01 \, n + 0.02 \, n^2 \,, \nonumber \\
    \delta_1(\lambda) &= -11.10 - 0.06 \, n  \,,\hspace{5mm}
    \delta_2(\lambda) = -0.04 - 0.03 \, n \,, \hspace{5mm}
    \delta_3(\lambda) = 5.62 + 0.87 \, n \,, \label{coeff_rho_rad}
\end{align}
and $n=-\log_{10}(\lambda)$. These expressions can be combined with the concept of \emph{heating efficiency} \cite{Rubio:2017gty}
\begin{equation} \label{eq:heating_eff}
    \Theta_{\text{ht}}(\lambda, \nu) \equiv \frac{\rho_\chi(z_{\rm rad})}{\rho_\phi(z_{\rm rad})} =\frac{\rho_{\rm rad}}{3 {H}_{\rm kin}^2 M_{P}^2}  \left(1+\frac{z_{\rm rad}}{\nu}\right)^3  \,,
\end{equation}
to gain a complete picture of the heating stage. In fact, this quantity will turn out to be very useful in the following discussion, as it simplifies the task of rescaling a GW signal to the present cosmological time. Lastly, the \emph{heating temperature} can be computed by associating an instantaneous temperature to the scalar field energy-density
\begin{equation} \label{eq:heating_temp}
T_{\rm ht} =\left(\frac{30\,\rho^{\chi}_{\rm ht}}{\pi^2 g_*^{\rm ht}}\right)^{1/4} \,.  
\end{equation}
where the subscript $ht$ indicates quantities at the time of a heating. Given that the typical heating and thermalisation timescales are of comparable duration \cite{Micha:2002ey, Micha:2003ws, Micha:2004bv}, the previous definition is a also good estimate of the actual heating temperature. We assume $g_*^{\rm ht}=106.75$ at the end of the actual heating phase.

\section{Stochastic background of gravitational waves} \label{sec:GW}

As the spectator field undergoes the tachyonically-unstable phase, its quantum fluctuations are exponentially amplified, leading to large inhomogeneities on patches with typical scales $\sim \mathcal{H}^{-1}$. These are the main large-amplitude source of the SGWB we will analyse through lattice simulations. In this Section, we describe the underlying physics and get an analytical understanding of the final GW spectrum. The complete set of assumptions and explicit derivation can be found in Appendix \ref{app:GW_source}. 

In the spectator field approximation ($\xi\chi^2\ll M_P^2$, $\rho_{\chi}\ll \rho_\phi$),\footnote{Notice that, in the limit of a spectator scalar field the correction to the Planck mass is negligibly small. In \cite{Bettoni:2019dcw, Laverda:2023uqv} for a field to be a \emph{spectator}, the non-minimal gravitational coupling is allowed to contribute up to $10\%$ of the Planck mass squared. This upper bound can be saturated only at the end of the first semi-oscillation for high values of non-minimal coupling parameter $\xi$, high kination scales $H_{\rm kin}$ and small values of self-coupling $\lambda$. The parameter space in which we will perform our analysis is limited to small values of $\xi$ that can at most cause a $1\%$ change in $M_P^2$ for a kination scale below $10^{12} \textrm{GeV}$ and even smaller changes for lower scales. For $\nu^6 H_{\rm kin}^2 \lesssim 0.5 \times M_P^2 \lambda$, our results are also compatible with the general restriction $\langle \chi^2\rangle^{1/2} \lesssim M_P/\xi $ imposed by the quantum self-consistency of this type of non-minimally scalar-tensor theories, see e.g. \cite{Rubio:2018ogq} and references therein. This condition is satisfied in particular for all figures displayed in this work, with $\langle \chi^2\rangle$ at backreaction time computed from the fitting formulas in \cite{Laverda:2023uqv}. Once this threshold is surpassed, the original Lagrangian density \eqref{eq:lagchi} should be potentially supplemented by an infinite series of higher-dimensional operators to preserve consistency, in line with the standard effective-field-theory approach.} the evolution equation for a transverse-traceless (TT) gravitational wave perturbation $h^{TT}_{ij}$ ($\partial_i h^{TT}_{ij}=h^{TT}_{ii}=0$) on a flat FLRW background 
\begin{equation}
    ds^2 = -dt^2+a^2(t)(\delta_{ij}+h^{TT}_{ij})dx^i dx^j\,,
\end{equation}
takes the form 
\begin{equation} \label{eq:eom_GWs}
    \ddot h_{ij}^{TT}+ 3H\dot h_{ij}^{TT} - \frac{\nabla^2 h_{ij}^{TT}}{a^2} \simeq \frac{2}{a^2M_P^2}  {\Pi}^{TT}_{ij}\,,
\end{equation}
with the source term ${\Pi}^{TT}_{ij}$ standing for the TT part of the effective anisotropic energy-momentum tensor
\begin{equation} \label{eq:TT_stress_energy_tensor}
    {\Pi}_{ij} = \partial_i\chi\partial_j\chi  - \xi\partial_i\partial_j\chi^2 \,.
\end{equation}
Knowing the solution of \eqref{eq:eom_GWs}, the energy-density in GWs can then be computed as the ensemble average
\begin{equation}\label{rhoGW}
    \rho_{\rm GW}= \frac{M_P^2}{4}\langle  \dot{h}^{TT}_{ij}(t,\mathbf{x}) \dot{h}^{TT}_{ij}(t,\mathbf{x}) \rangle \;, 
\end{equation}
with the corresponding power spectrum per logarithmic frequency interval given by
\begin{equation} \label{eq:omega_GW_definition}
    \Omega_{\rm GW}=\frac{1}{\rho_c}\frac{d \rho_{\rm GW}}{d \log k}\,,
\end{equation}
where $\rho_c$ normalises the GW spectrum to a critical energy-density. We expect the GW spectrum to be peaked at momenta somewhat above $H_{\rm kin}$, as the backreaction associated to the quartic self-interaction typically breaks down the larger homogeneous patches into sub-horizon fluctuations. The brief life of these colliding ``bubbly" features in the field distribution sources a peaked GW signal \cite{Garcia-Bellido:2007fiu}. Indeed, there exist a certain similarity to the generation of GWs in first-order phase transitions, since the large-amplitude fluctuations are coherent on Hubble-sized patches. One can now clearly appreciate the need for lattice simulations, as such spatial non-linear features can only be studied in a fully-interacting system. Any homogeneous approximation of the dynamics \cite{Opferkuch:2019zbd, Dimopoulos:2018wfg, Figueroa:2016dsc, Fairbairn:2018bsw} is indeed not suitable for achieving a consistent physical description of the process \cite{Felder:2001kt}. 

\section{Non-linear dynamics and lattice simulations} \label{sec:lattice}

Studying the production of GWs during the non-linear phase that follows the onset of backreaction effects requires the use of classical lattice simulations in $3+1$ dimensions. To this end, we will make use of a modified version of the public code \texttt{$\mathcal{C}osmo\mathcal{L}attice$} \cite{Figueroa:2020rrl, Figueroa:2021yhd} to evolve a self-interacting scalar field in an expanding background while solving the tensor perturbation evolution equation \eqref{eq:eom_GWs} that will give us the GW power spectrum \eqref{eq:omega_GW_definition}. The default lattice implementation of the Klein-Gordon equation of the spectator field has been modified according to \eqref{eq:klein_gordon_spect} to account for the Hubble-dependent mass term. The same is true for the energy momentum tensor in \eqref{eq:energy_momentum_tensor} and the total energy and pressure \eqref{eq:energy_density_pressure} which include additional terms derived from the extra non-minimal interaction with gravity. The TT stress-energy tensor that sources GWs has also been modified according to the expression in \eqref{eq:TT_stress_energy_tensor}. We follow some uniform prescriptions in all our lattice simulations to ensure the stability and reliability of the output:
\begin{enumerate} 

    \item  The number of lattice points per dimension $N=288$ (and therefore the comoving lattice size $L$) are selected in such a way that all relevant modes are always well within the associated infrared (IR) and ultraviolet (UV) resolution in momentum space, thus covering the tachyonic band \cite{Bettoni:2018utf} as well as modes enhanced by subsequent rescattering effects. This leads us to setting $\kappa_{\text{IR}} = 2\nu \mathcal{H}(0) $ and $\kappa_{\text{UV}} \gg  \sqrt{4\nu^2 - 1}\kappa_{\rm IR}$ with $\kappa = k / H_{\rm kin}$. 

    \item The time-step variable is chosen according to the stability criterion $\delta t / \delta x \ll 1/\sqrt{d}$ \cite{Figueroa:2021yhd}, with $d=3$ the number of spatial dimensions and $\delta x = 2\pi/(N \kappa_{\rm IR})$ the lattice spacing. More specifically,  we set $\delta t=0.1$ for $\nu \geq 10$ and $\delta t=0.01$ for $\nu < 10$. 

    \item The background expansion in \eqref{eq:backkin} is given by a fixed power-law and therefore, we adopt a symplectic 4th order Velocity-Verlet evolver to achieve a satisfactory stability and precision of the numerical solutions when the conservation of energy cannot be explicitly checked. 

    \item The lattice initial conditions are set as $\chi(0)=\chi'(0)=0$, as the spectator quantum field finds itself in vacuum at the end of inflation, with fluctuations over the homogeneous background included as Gaussian random fields. Given the fast classicalisation of the field's quantum fluctuations, the resulting evolution is deterministic up to a randomly-generated initial seed. We choose to keep this base seed constant in all our simulations so to make them exactly comparable, although a more robust but time-consuming approach would require repeating the same simulations with different random seeds and averaging over them. Given that the system looses memory of the initial conditions soon after the development of the tachyonic instability, fixing the initial random seed does not influence the overall macroscopic lattice-averaged evolution. 
\end{enumerate}

\begin{figure}[tb]
\centering
    \includegraphics[width=0.47\textwidth]{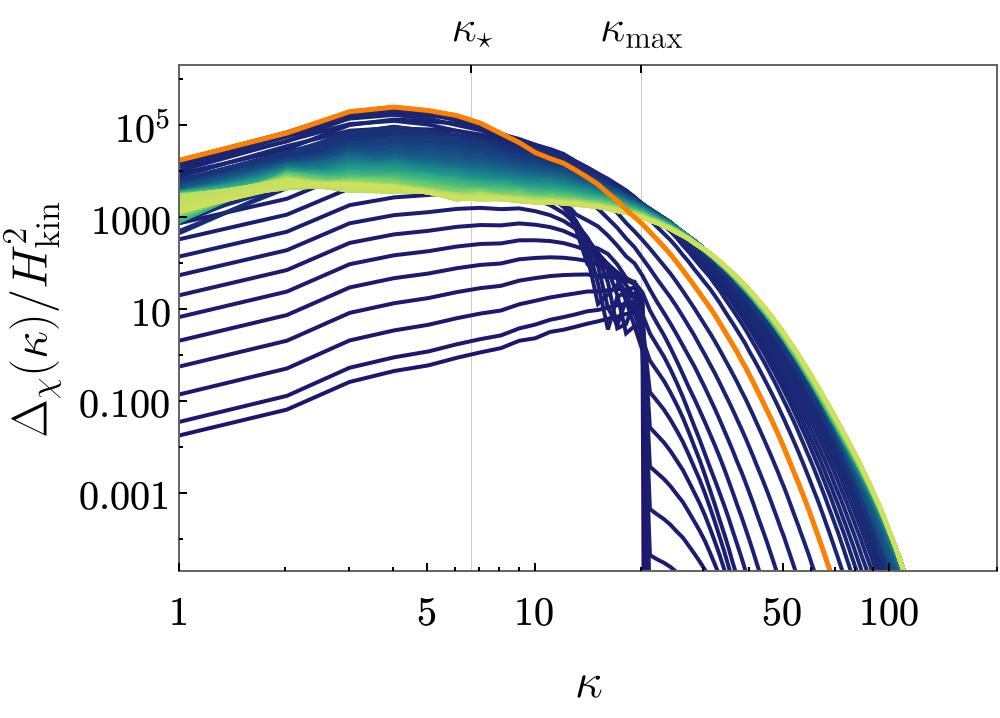}
    \hspace{5mm}
    \includegraphics[width=0.47\textwidth]{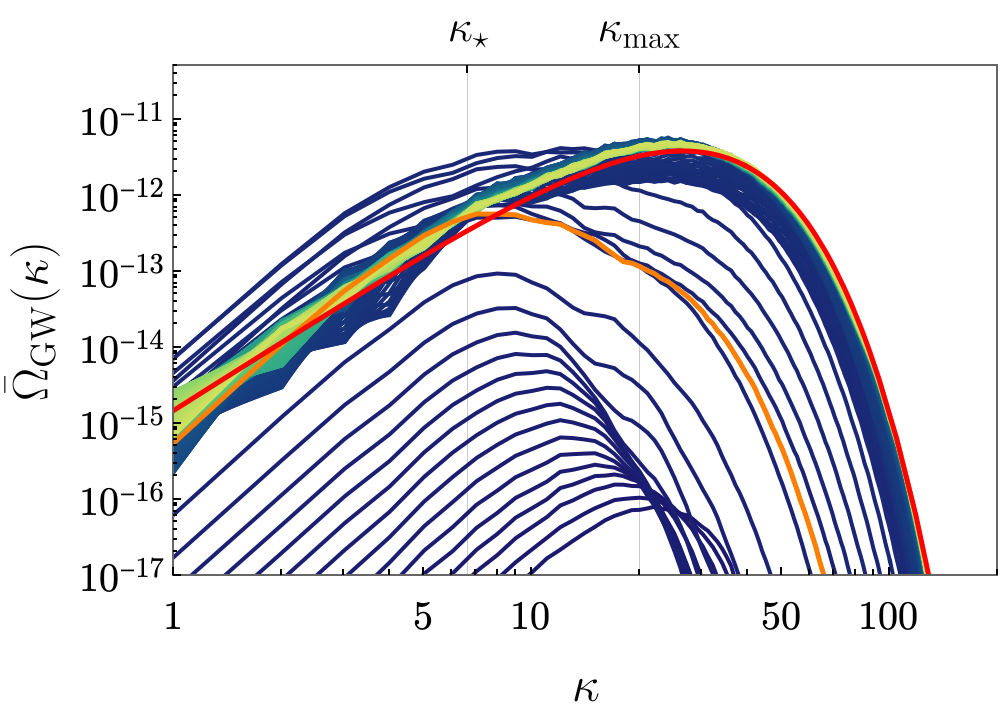}
    \caption{Typical output of a lattice simulation with $\nu=10$, $\lambda=10^{-4}$ and $H_{\rm kin}=10^{10} \textrm{ GeV}$ showing the power spectrum $\Delta_{\chi}(\kappa)$ of the spectator field and the GW spectrum $\Bar\Omega_{\rm GW}(\kappa)$ normalised with respect to the total energy density of the spectator field. Grey vertical lines indicate the momenta scales corresponding to the typical amplified momentum $\kappa_{\star}$ and the maximum amplified momentum $\kappa_{\rm max}$ in the tachyonic band at $z=0$. The spectra are plotted from the initial time until $z_{\rm rad}$ is reached. The orange line corresponds to the spectra measured at $z_{\rm br}$ while the red line in the second panel shows the fitted spectral shape in \eqref{eq:spectral_shape}.}
    \label{fig:spectra}
\end{figure}

The typical output of a lattice simulation is displayed in Figure \ref{fig:spectra} for the model parameters $\nu=10$, $\lambda=10^{-4}$. The scalar power spectrum $\Delta_{\chi}(k)=k^3\mathcal{P}_{\chi}/(2\pi)$ is shown in the first plot, where the two-point correlator is defined as $\langle \chi_{\mathbf{k}} \chi_{\mathbf{k}'}\rangle = (2\pi) \mathcal{P}_{\chi} \delta(\mathbf{k}-\mathbf{k}')$, while the GW power spectrum $\Bar\Omega_{\rm GW}(\kappa)$ is shown in the second plot. Notice that a bar on top of $\Bar\Omega_{\rm GW}$ indicates that it is normalised with respect to the total energy density of the spectator field, i.e. $\rho_c=|\rho_\chi|$.~\footnote{As the non-minimal interaction dominates the early dynamics, the total energy density of the field $\chi$ is not positive definite, hence the need for taking the absolute value \cite{Laverda:2023uqv, Bettoni:2021qfs, Bettoni:2021zhq}. At later times, as the non-minimal coupling becomes less relevant, $\rho_\chi$ becomes positive. In any case, the overall energy budget of the Universe is always a positive quantity \cite{Ford:1987de, Ford:2000xg,Bekenstein:1975ww,Flanagan:1996gw}.} The typical tachyonic momentum scale $\kappa_*= 2\sqrt{\nu+1}\, \kappa_{\rm IR}$ and the maximum momentum in the tachyonic band $\kappa_{\rm max}= \sqrt{4\nu^2 - 1}\kappa_{\rm IR}$ are highlighted. Some features are to be noted. After the initial violent tachyonic phase, the system slowly relaxes towards an equilibrium state with a steady flow of energy from IR modes to UV ones. The production of GWs is peaked around the time of backreaction $z_{\rm br}=13$ for the chosen model parameters, i.e., when the tachyonically-amplified fluctuations reach their maximum extension and lead to the largest contribution to the anisotropy of the system. The orange lines in both plots highlight the spectra at $z_{\rm br}$. After the large-amplitudes oscillations have been quenched, GWs are produced at a lower rate and the spectrum tends to achieve a steady configuration, with some residual sources given by the small-scale spatial oscillations of the spectator field. The final shape at $z_{\rm rad}$ is given by a simple broken-power law, highlighted by a red line in the figure.    

We perform an extensive scanning of the parameter space that covers a large range in non-minimal coupling parameters $\nu \in [6,\,24]$, equivalent to $\xi \in [24, \, 384]$, thus extending the most typically studied range \cite{Figueroa:2016dsc} up towards Higgs-Inflation-like values \cite{Bezrukov:2007ep, Rubio:2018ogq}. Regarding the self-interaction parameter, we probe the range $\lambda \in [10^{-1}, \, 10^{-6}]$ which covers, among other scenarios, the Standard-Model running of the Higgs self-coupling up to the proximity of the instability scale \cite{Bezrukov:2012sa,Degrassi:2012ry,Buttazzo:2013uya,Bezrukov:2014ipa,Rubio:2015zia}. Being the output of all simulations normalised with respect to the mass-scale $H_{\rm kin}$, each simulation is independent of the choice of such scale, which can be reintroduced whenever dimensionful quantities are needed.  

\section{Parametric fitting formulas and spectral shape} \label{sec:fitting_formulas}

\begin{figure}[tb]
\centering
\includegraphics[width=0.8\textwidth]{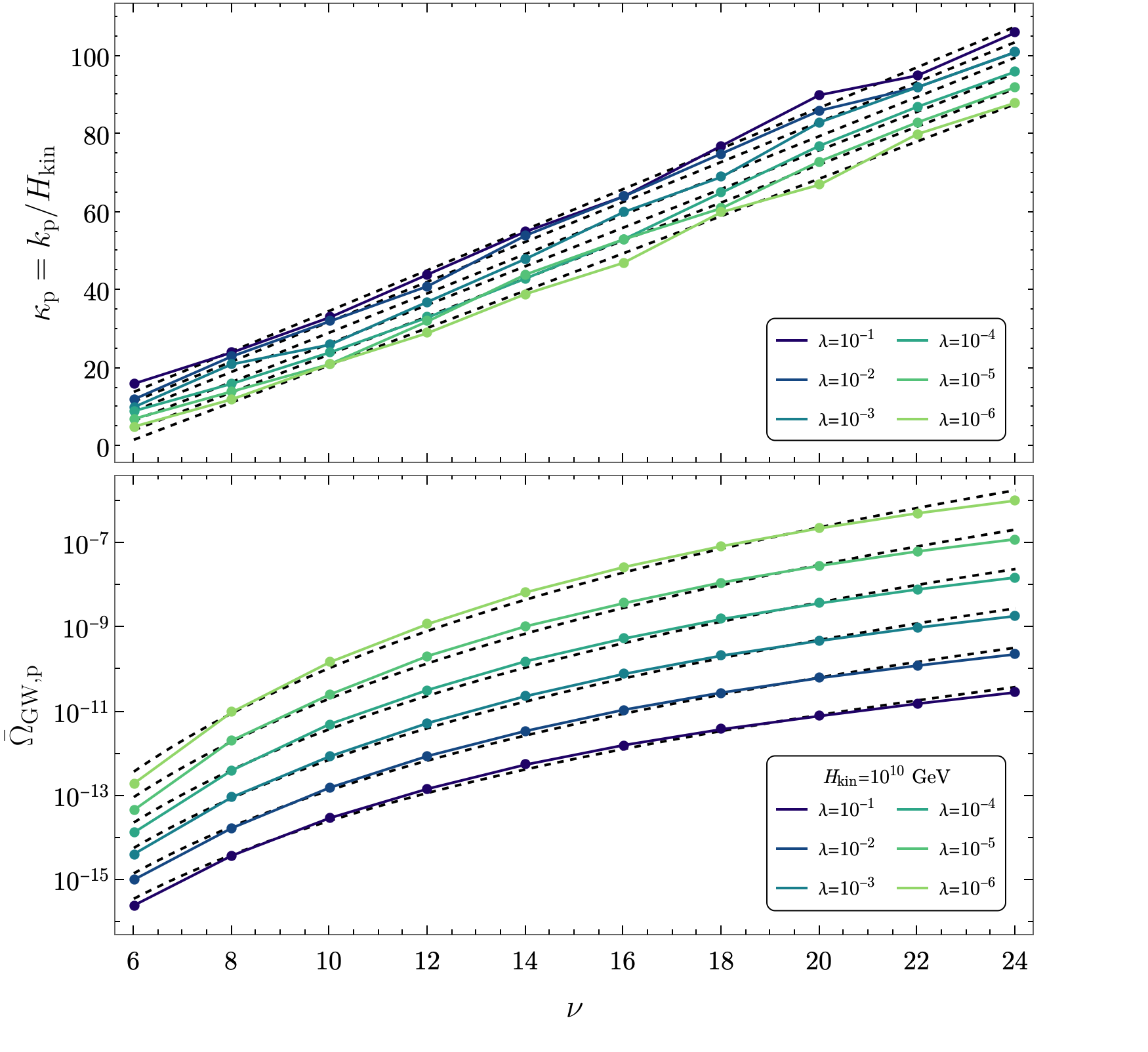}
\caption{Peak frequency $\kappa_{p}$ and peak amplitude $\Bar\Omega_{\rm GW,p}$ evaluated at $z_{\rm rad}$ as a function of the model parameters $(\nu, \, \lambda)$. Regarding $\Bar\Omega_{\rm GW,p}$, we have set a fiducial value of $H_{\rm kin}=10^{10} \textrm{ GeV}$. Dots indicate the output of single simulations in our scanning grid while black dashed lines display the resulting fitting formulas.} \label{fig:fitting_total}
\end{figure}

The scanning of the $(\nu, \, \lambda)$ parameter space allows us to condense a large amount of numerical information into a set of simple and useful fitting formulas, in the spirit of what has been previously done in the HIPT scenario \cite{Laverda:2023uqv} and for parametric-resonant reheating \cite{Figueroa:2016wxr}. We take the \emph{radiation time} $z_{\rm rad}$ as the typical timescale at which we analyse the GW spectrum in each simulation. As was shown in \cite{Laverda:2023uqv}, this timescale represents the moment in which the system gains its macroscopic lattice-averaged scaling behaviour, since the for $z>z_{\rm rad}$ the lattice-averaged scalar field equation of state is $\langle w_\chi \rangle \simeq 1/3$. We focus on measuring the peak frequency, peak amplitude and integrated energy density of the GW spectrum at $z_{\rm rad}$. Given that the energy-density stored into GWs also scales as radiation, it is convenient to use the total energy of the spectator field $\chi$ as the critical energy-density in \eqref{eq:omega_GW_definition} and define
\begin{equation}
    \Bar\Omega_{\rm GW}(z_{\rm rad})\equiv\frac{\rho_{\rm GW}(z_{\rm rad})}{\rho_{\chi}(z_{\rm rad})} \,.
\end{equation}
As most of the GW production process is active in the non-linear phase between backreaction $z_{\rm br}$ and $z_{\rm rad}$, measuring $\Bar\Omega_{\rm GW}(z_{\rm rad})$ gives a good estimate of the final GW energy density at the time of reheating. On top of the previous considerations, quantities at $z_{\rm rad}$ can be more straightforwardly used together with the parametric formulas in \cite{Laverda:2023uqv} which are also evaluated at the same timescale.  

Following the previous considerations, we compute two fitting formulas from the GW spectra at $z_{\rm rad}$ for the peak momentum $\kappa_{\rm p}$ and the peak amplitude ${\Bar\Omega_{\rm GW, p}=\rho_{\rm GW, p} / |\rho^{\chi}|}$ as a function of $\nu$, $n=-\log_{10} \lambda$ and $H_{\rm kin}$. For simplicity's sake, we always understand the quantities $\kappa_{\rm p}$, ${\Bar\Omega_{\rm GW, p}}$ and ${\Bar\Omega_{\rm GW}}$ to be measured at $z_{\rm rad}$ without writing it explicitly for the remainder of this work. Using simple linear and exponential ansatz, we obtain that the peak momentum is given by
\begin{gather} \label{eq:fit_momentum_peak}
    \kappa_{\rm p}(\nu, \lambda) = \alpha_1 + \alpha_2 \nu \; , \hspace{3mm} \alpha_1 = -15.36 - 1.95 n \; , \hspace{3mm} \alpha_2 = 5.29 - 0.08 n \; ,
\end{gather}
while the peak amplitude is
\begin{gather} \label{eq:fit_omega_peak}
    \Bar\Omega_{\rm GW, p}(\nu, \lambda) = \left(\frac{H_{\rm kin}}{10^{10} \textrm{ GeV}}\right)^2 \exp \left[ \beta_1 + \beta_2 \log \nu \right] \; , \nonumber \\ 
    \beta_1 = -50.92 + 0.40 n \; , \hspace{5mm}
    \beta_2 = 7.80 + 0.55 n \; .
\end{gather}
A formula for the integrated energy density of GWs can be obtained in the same way and, as can be expected from peaked production processes, it matches closely the expression for $\Bar\Omega_{\rm GW, p}(\nu, \lambda)$, namely 
\begin{equation} \label{eq:fit_omega_int}
\begin{aligned} 
   & \Bar\Omega_{\rm GW}(\nu, \lambda) = \left(\frac{H_{\rm kin}}{10^{10} \textrm{ GeV}}\right)^2 \exp \left[ \gamma_1 + \gamma_2 \log \nu \right] \; , \\
   & \gamma_1 = -50.50 + 0.40 n \; , \hspace{5mm}
    \gamma_2 = 7.70 + 0.55 n \; ,
\end{aligned}
\end{equation}
where the factor of $(H_{\rm kin}/10^{10} \textrm{ GeV})^2$ comes from the normalisation of tensor perturbations in the lattice code \cite{cosmolattice_technical_notes}. The lattice-derived data points and the respective fitting functions can be seen in Figure \ref{fig:fitting_total}. We note here that the fitting procedure has been monitored using a $R^2$-test that always yielded values greater than $0.99$. As one would expect, a more intense tachyonic phase leads to a larger portion of energy to be transferred to GWs, as is the case for large values of $\nu$ and small values of $\lambda$. At the same time, the spectrum is typically peaked at momenta $\kappa_{\rm p}$ one or two orders of magnitude larger than the Hubble scale at kination. This is an expected feature since the initial IR amplification breaks turbulently into UV inhomogeneities while GWs are produced. 

\begin{figure}[tb]
\centering
    \includegraphics[width=0.45\textwidth]{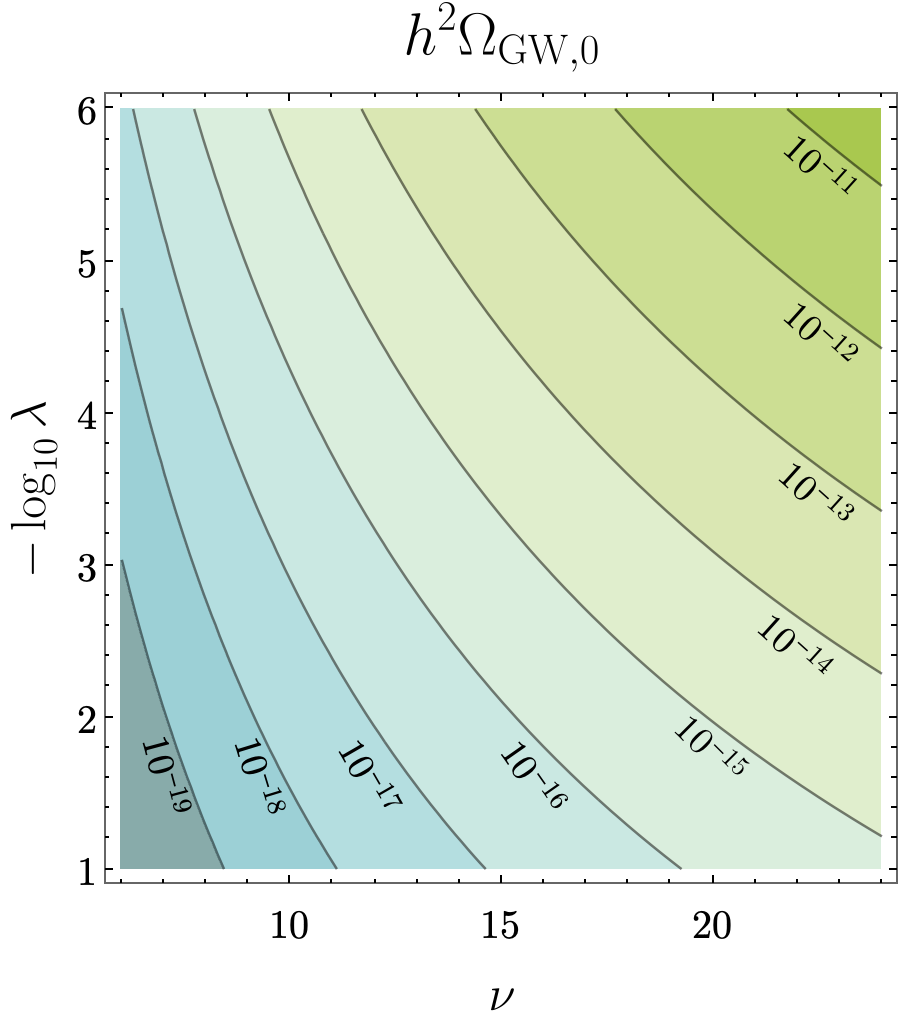}
    \hspace{5mm}
    \includegraphics[width=0.45\textwidth]{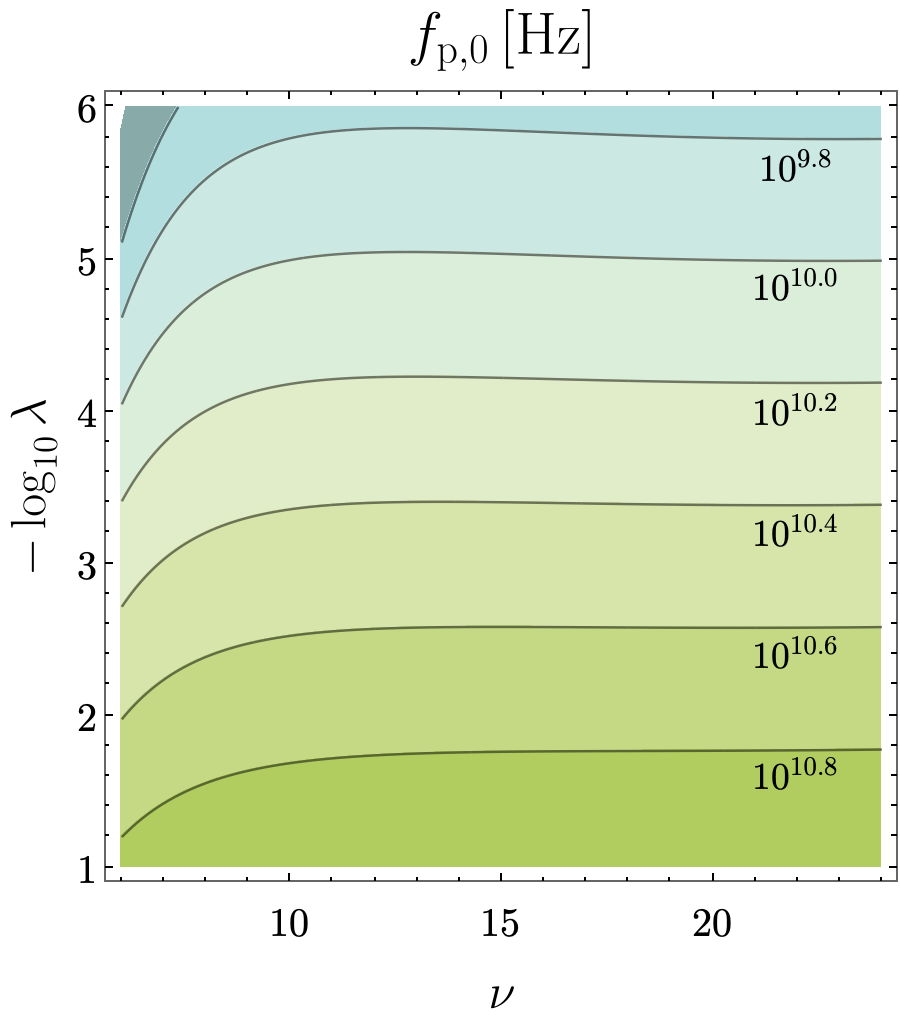}
    \hspace{5mm}
    \caption{Integrated GW energy density $h^2\Omega_{\rm GW, 0}$ and frequency of the peak $f_{p,0}$ at the present cosmological time as a function of the model parameters $\nu$ and $\lambda$ within the parameter space under study. For the energy-density contours, we have set a fiducial value $H_{\rm kin}=10^{10} \textrm{ GeV}$. }
    \label{fig:max_energy_frequency_now}
\end{figure}

Once we have a complete picture of the GW spectrum at $z_{\rm rad}$, frequencies and energy-densities can be rescaled to the present time taking into account the full cosmic history post-production. Following the prescriptions in the literature \cite{Allahverdi:2020bys, Cosme:2022htl, Caprini:2018mtu}, we obtain
\begin{equation} \label{eq:omega_GW_0}
    \Omega_{\rm GW,0} = 1.67 \times 10^{-5} h^{-2} \left( \frac{100}{g_*^{\rm ht}} \right)^{1/3} \times \Bar{\Omega}_{\rm GW}\,,
\end{equation}
Indeed, $\Bar\Omega_{\rm GW}=\rho_{\rm GW} / |\rho^{\chi}|$ measures already the energy-density of GWs at the end of the heating phase, when the spectator field becomes the dominating energy component of the early Universe. The numerical factor encodes the usual dilution due to matter-domination. The dependence of this quantity on the scale of kination is only present implicitly in the fitting formula \eqref{eq:fit_omega_int}. Regarding the shifting in frequency of the spectrum, we compute the rescaling to be
\begin{equation}\label{eq:f_0}
    f_{p,0} = \frac{a_{\rm rad}}{a_0} f_{p,\rm rad} \simeq 1.3\times 10^9 \, \text{Hz} \; \frac{\kappa}{2\pi} \left(\frac{H_{\rm kin}}{10^{10} \textrm{ GeV}}\right)^{1/2} \left( \frac{\Theta_{\rm ht}}{10^{-8}}\right)^{-1/4} \sqrt{a_{\rm rad}}\,,
\end{equation}
where we have normalised the scale factor to the kination scale $a_{\rm kin}=1$, set $a_{\rm eq}/a_0 \simeq 1/3400$, $H_{\rm eq}\simeq 10^{-37} \text{ GeV}$ and identified $1 \text{ GeV}=1.5 \times 10^{24} \text{Hz}$. We have also used the relations
\begin{equation}
    \Theta_{\rm ht}=\left(\frac{a_{\rm ht}}{a_{\rm rad}}\right)^{-2}=\left( \frac{H_{\rm ht}}{H_{\rm rad}}\right)^{2/3} \; ,
    \hspace{4mm} \left(\frac{a_{\rm ht}}{a_{\rm eq}}\right)=\left(\frac{H_{\rm eq}}{H_{\rm ht}}\right)^{1/2} \; ,
    \hspace{4mm} \left(\frac{a_{\rm rad}}{a_{\rm ht}}\right)=\left(\frac{H_{\rm ht}}{H_{\rm rad}}\right)^{1/3} \; .
\end{equation}
involving the different energy scalings during kination and radiation-domination. The tachyonic production process leads to a typical energy density of the scalar field $\rho_\chi \sim H_{\rm kin}^4$, which translates into to $\Theta_{\rm ht} \sim H_{\rm kin}^2$. Indeed, in terms of the parametric formulas in \cite{Laverda:2023uqv}, we have 
\begin{equation}
    \Theta_{\text{ht}}(\lambda, \nu) = \frac{16}{3}\left(\frac{H_{\rm kin}}{M_{P}}\right)^2  \left(1+\frac{\gamma_1 + \gamma_2 \, \nu}{\nu}\right)^3 \exp\left({\delta_1 + \delta_2 \, \nu} +\delta_3 \ln \nu \right)\,,
\end{equation}
and all dependence on the kination scale in \eqref{eq:f_0} vanishes. This exact cancellation is an effect of the tight connection between $H_{\rm kin}$ and the scale of heating $H_{\rm ht}$ via the heating efficiency. Qualitatively speaking, the period of kination causes a blueshift of the spectra frequencies that is $H_{\rm kin}$-dependent and compensates exactly any change in the initial scale $H_{\rm kin}$. Figure \ref{fig:max_energy_frequency_now} shows the parametric dependence of $f_{\rm p,0}$ and $\Omega_{\rm GW,0}$, the latter for an indicative scale of kination of $H_{\rm kin}=10^{10} \textrm{ GeV}$. The typical GW spectrum in our chosen parameter space is placed at $10-100 \textrm{ GHz}$, while its amplitude can be as large as $\mathcal{O}(10^{-10})$ for $H_{\rm kin}=10^{10} \textrm{ GeV}$. It is interesting to notice that lower values of $\lambda$ cause the spectrum to be shifted at smaller frequencies.   

\begin{figure}[tb]
\centering
\includegraphics[width=0.8\textwidth]{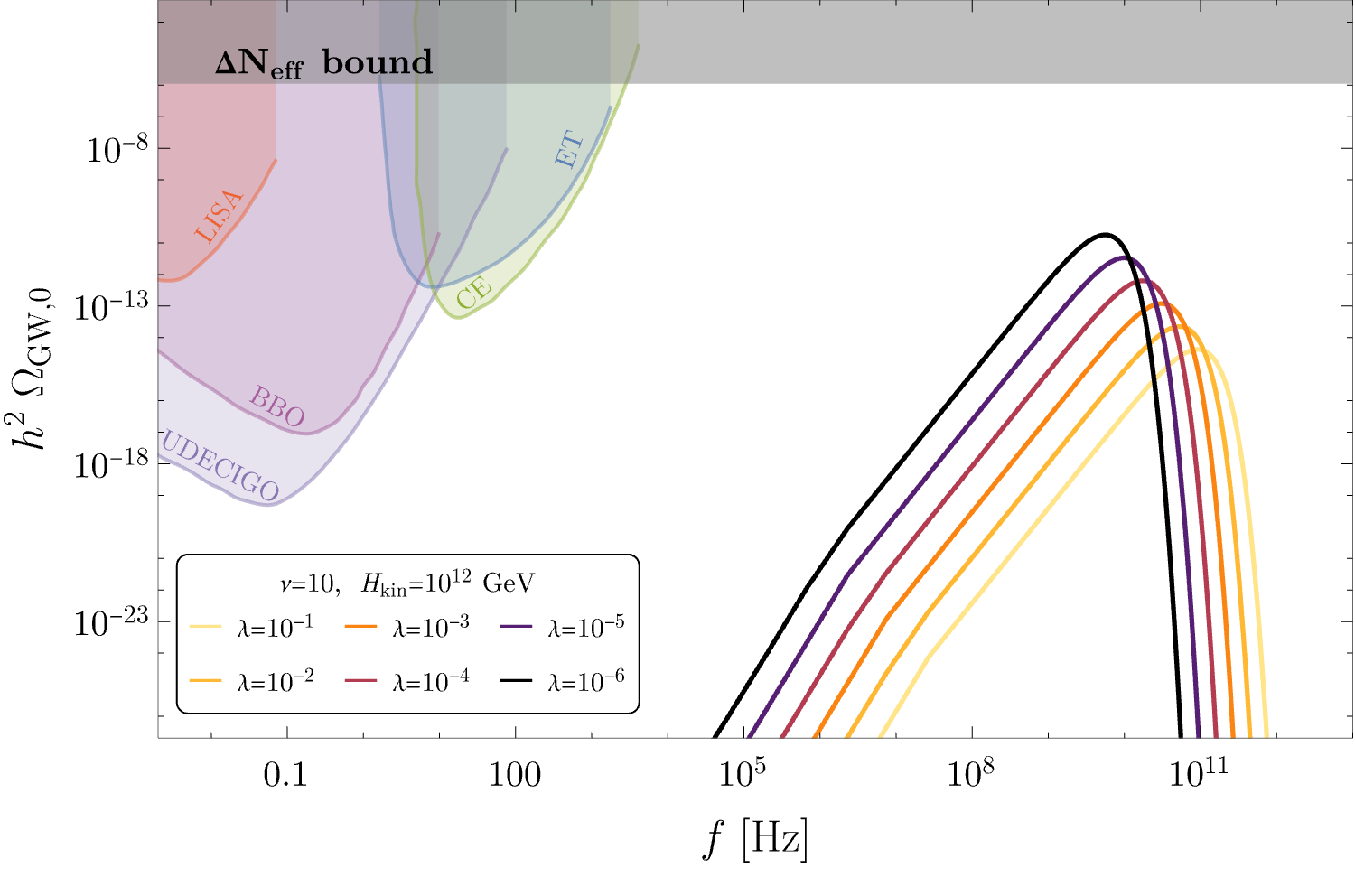}
\hspace{7mm}
\caption{Spectra of the SGWB signal from a HIPT as a function of the self-coupling parameter $\lambda$ for a fixed value of $\nu=10$ and $H_{\rm kin}=10^{12} \textrm{ GeV}$. Several sensitivity curves of proposed future detectors are being shown:  Laser Interferometer Space Antenna (LISA) \cite{amaroseoane2017laserinterferometerspaceantenna, Robson:2018ifk}, Big Bang Observer (BBO) \cite{Crowder:2005nr, Corbin:2005ny}, UltimateDECIGO \cite{Seto:2001qf, Yagi:2011wg, Kawamura:2020pcg}, Einstein Telescope (ET) \cite{Punturo:2010zz, Branchesi:2023mws}, and Cosmic Explorer (CE) \cite{LIGOScientific:2016wof, Reitze:2019iox}. The grey-shaded area corresponds to the region excluded by the bound on the integrated energy-density in GWs at BBN in \eqref{eq:Neff_bound}. } \label{fig:GW_detectors_lambda}
\end{figure}

Given the extensive information about GW spectra at $z_{\rm rad}$ obtained previously, it is possible to match the simulation output to a simple broken power-law fit. A customary ansatz for peaked sources of GWs is given by the shape function 
\begin{equation} \label{eq:spectral_shape}
    \Bar{\Omega}_{\rm GW}(f)=\frac{\Bar{\Omega}_{\rm GW,p}(a+b)^c}{\left[a \left(\frac{f}{f_{\rm p}}\right)^{b/c}+b \left(\frac{f}{f_{\rm p}}\right)^{-a/c}\right]^c}\,,
\end{equation}
frequently adopted to describe the GW signal from first-order phase transitions \cite{Lewicki:2020azd, Lewicki:2022pdb}. The fitting procedure consists in computing the best values for free parameter $a$, $b$ and $c$ in \eqref{eq:spectral_shape} for each simulation output using a $\chi$-squared algorithm. The coefficients $a$ and $b$ control the IR and UV slopes respectively, while $c$ determines the overall width. Spectra taken at $z_{\rm rad}$ are typically self-similar and smooth, as the production process has already quenched. This fact allows us to simply perform the fitting procedure for every power spectrum extracted at $z_{\rm rad}$ and compute the coefficient's best values, finding
\begin{equation} \label{eq:fitting_parameters_abc}
    a=3.00\,, \qquad b = 152.34 - 6.57 \, \nu \,, \qquad c=105.85 - 4.79 \, \nu \;.
\end{equation}
The average value $a=3$ is expected from causality arguments at small momenta \cite{Caprini:2009fx, Cai:2019cdl}, while both $b$ and $c$ show a decreasing trend proportional to $\nu$. Such a feature is linked to the underlying differences of the system at $z_{\rm rad}$: this timescale is generically longer for more intense tachyonic phases (large $\nu$), thus causing the appearance of larger fluctuations in the UV region. Therefore, the resulting GW spectrum is less steep in the UV as compared to weaker tachyonic phases (small $\nu$).

It has to be noted that two factors can contribute to the uncertainty in measuring the shape coefficients. In the IR region, because of the strong initial amplification of the scalar field modes, some secondary peaks can still be present at the time $z_{\rm rad}$. These lead to the appearance of a somewhat flatter portion of the spectrum around $\kappa_\star$, with a $\sim f^2$ scaling (see Figure \ref{fig:spectra}). In the UV region, the finite size of the lattice and its limited capability in resolving small modes can lead to some well-known artefacts that can be avoided with an optimal choice of lattice parameters \cite{Cui:2023fbg, Dankovsky:2024zvs}. By sampling the parameter space with a few simulations with increased lattice size, we have verified that $N=288$ is an adequate number of lattice points to avoid the formation of such artifacts while covering all of the relevant momenta range. Moreover, different definitions of power spectra can lead to discrepancies in the UV counting of modes \cite{cosmolattice_technical_notes}. In order to minimise the effects of such unphysical artifacts, we set the fitting domain to be $\kappa<\kappa_{\rm UV}$ and we examine only the spectra of those simulations with $10<\nu<20$. Following this approach, the values in \eqref{eq:fitting_parameters_abc} give a good estimate of the overall shape of the spectrum and can be adopted to study a larger range of frequencies. Notice also that the IR and UV features we are discussing have a negligible effect on the integrated power spectrum.

\begin{figure}[tb]
\centering
\includegraphics[width=0.8\textwidth]{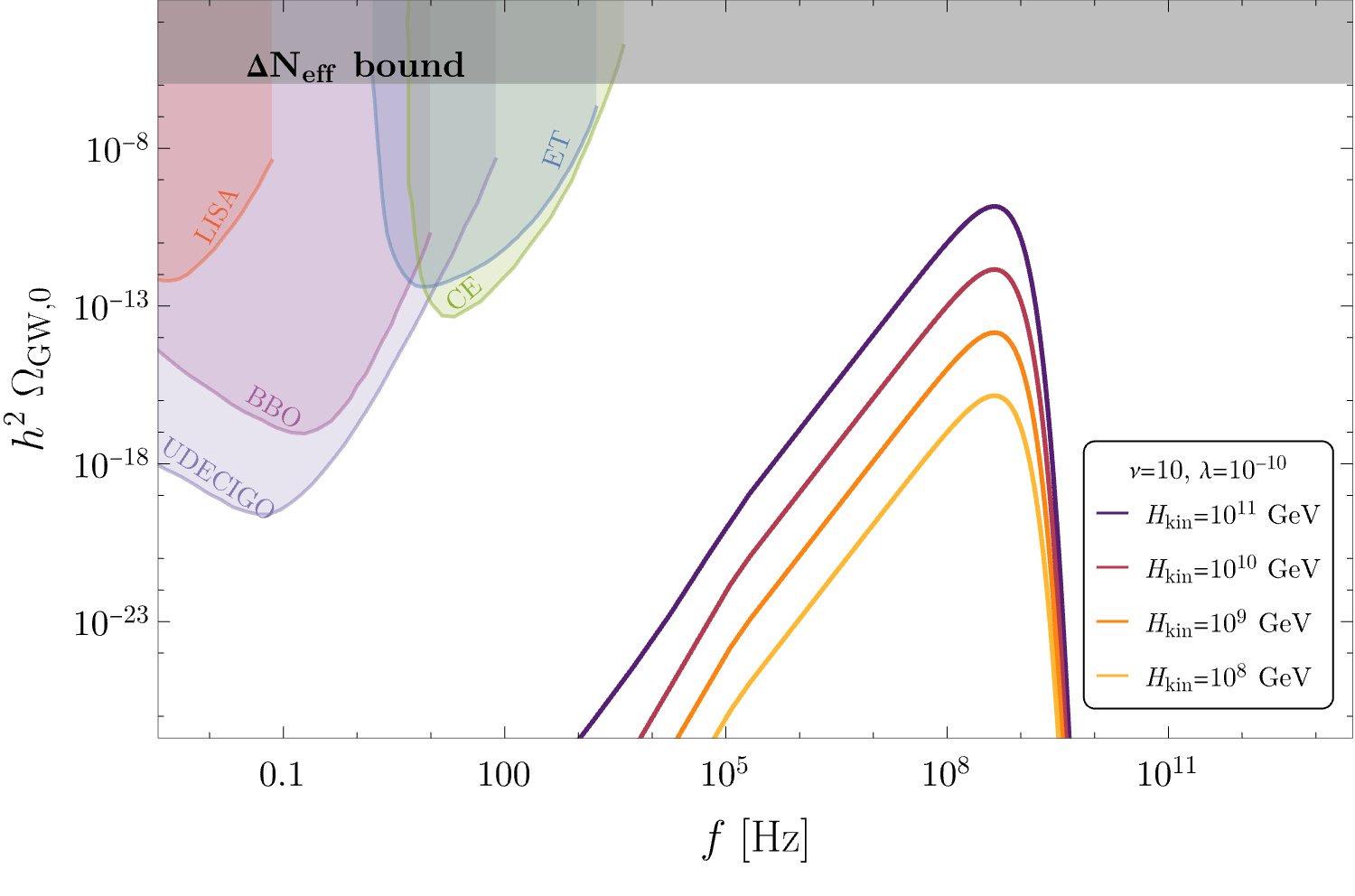}
\hspace{7mm}
\caption{Spectra of the SGWB signal from a HIPT as a function of the scale of kination $H_{\rm kin}$ for fixed values of $\nu=10$ and $\lambda=10^{-10}$. The grey-shaded area corresponds to the region excluded by the bound on the integrated energy-density in GWs at BBN in \eqref{eq:Neff_bound}.}\label{fig:GW_detectors_hkin}
\end{figure}

\section{Potential for detection} \label{sec:detectors}

The final step in our analysis consists in comparing the fitted spectra with the sensitivity curves of various proposed GW detectors. Figure \ref{fig:GW_detectors_lambda} shows the spectra sourced by the HIPT, rescaled to the present time and for different values of $\lambda$ and a fiducial scale of kination $H_{\rm kin}=10^{12} \textrm{ GeV}$. The GW signal produced by the phase transition is always fulfilling the BBN bound on the relativistic number of degrees of freedom $\Delta N_{\rm eff}$ for all the parameter space under consideration \cite{Caprini:2018mtu}. Indeed, any integrated contribution of a GW power-spectrum has to satisfy the constraint \cite{Planck:2018vyg}
\begin{equation} \label{eq:Neff_bound}
    h^2 \int \frac{df}{f} \Omega_{\rm GW,0} \lesssim  5.6 \times 10^{-6} \Delta N_{\rm eff} = 1.1 \times 10^{-6}\,,
\end{equation}
which can be recast as a condition on the parameters of the model, namely
\begin{equation} \label{eq:Neff_bound_param}
    \Bar\Omega_{\rm GW}(\nu,\lambda,H_{\rm kin})\lesssim 6.7 \times 10^{-2} \; .
\end{equation}
For $H_{\rm kin}=10^{12} \textrm{ GeV}$, the bound excludes a portion of parameter space outside our chosen parameter range, see the grey-shaded region in Figure \ref{fig:GW_detectors_lambda}. The tilt in the low-frequency portion of the spectrum depends on the expansion rate at the horizon-reentry of IR modes, namely $\Omega_{\rm GW,0}\sim f^4$ during kination and $\Omega_{\rm GW,0}\sim f^3$ during radiation-domination \cite{Gouttenoire:2021jhk, Hook:2020phx}.

\begin{figure}[tb]
\centering
\includegraphics[width=0.8\textwidth]{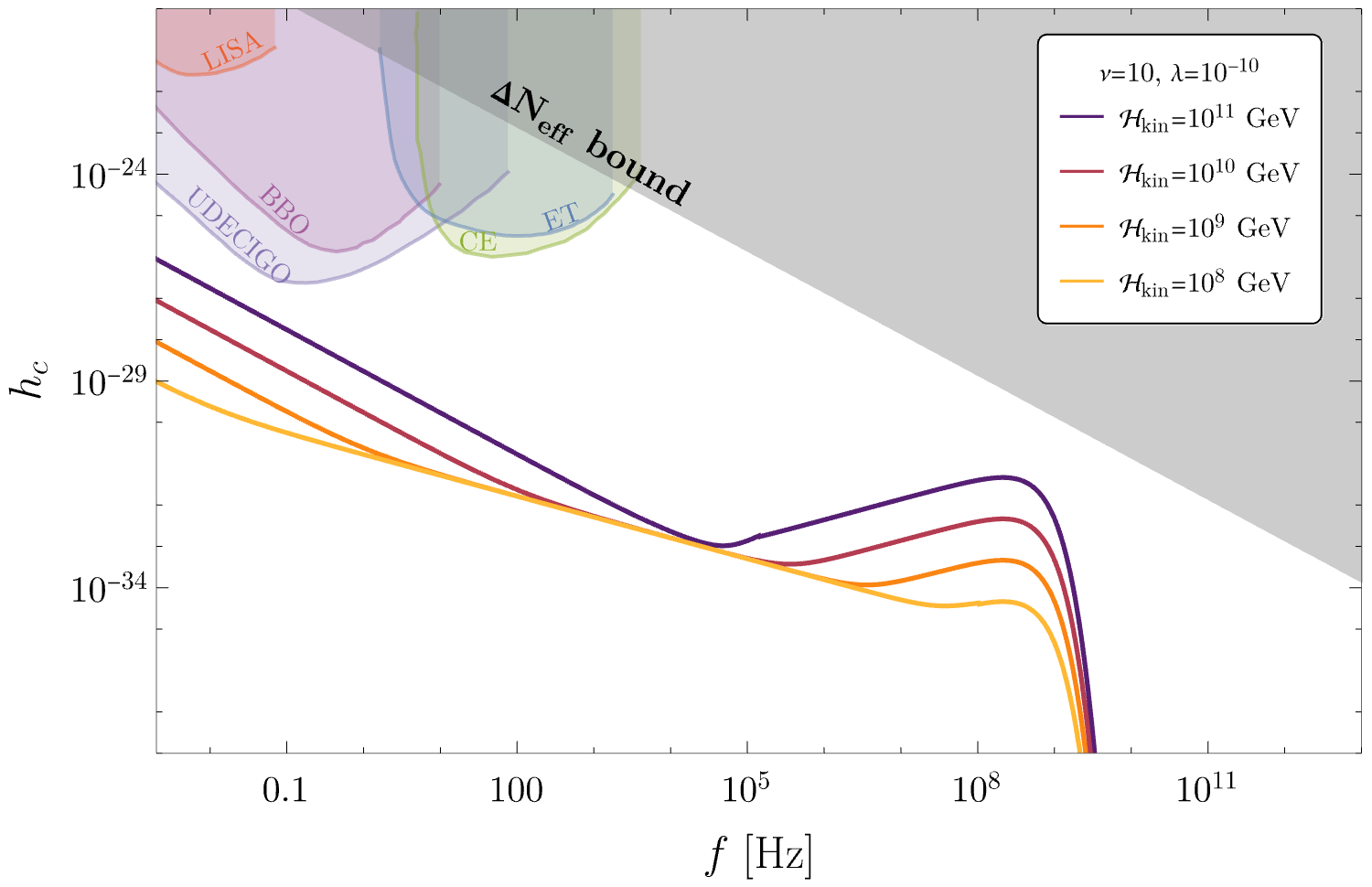}
\hspace{7mm}
\caption{Stochastic strain $h_c$ of the GW background from a HIPT for different kination scales while the other model parameters are set to $\lambda=10^{-10}$ and $\nu=10$. The bound on the effective numbers of degrees of freedom at BBN is shown as a grey-shaded area.} \label{fig:GW_detectors_strain}
\end{figure}

As noted in Figure \ref{fig:max_energy_frequency_now}, the larger anisotropy generated by stronger tachyonic phases sources a stronger signal at lower frequencies. This is the case for a lower value of $\lambda$, which drives the typical frequencies closer to the Hubble scale at the beginning of kination, as the subsequent heating phase is short-lived and radiation-domination is quickly achieved. This effect reduces the blueshifting, thus opening the possibility for peak detection at $10^8-10^9 \textrm{ Hz}$ as can be seen in Figure \ref{fig:GW_detectors_hkin}. Notice that in Figure \ref{fig:GW_detectors_hkin} we have fixed the scalar field self-coupling to be $\lambda=10^{-10}$ thus extrapolating from the clear trend in Figure \ref{fig:max_energy_frequency_now}, based on the simplicity and self-similarity of the fitting functions. 

In principle, the signal in Figure \ref{fig:GW_detectors_lambda} can be directly amplified by choosing larger kination scales due to the multiplication in \eqref{eq:fit_omega_int}. However, one has to take into account the existence of a maximum and minimum kination scale. The minimum scale guarantees reheating before BBN for a chosen set of model parameters, namely that $T_{\rm ht}\gtrsim 5 \textrm{ MeV}$ \cite{deSalas:2015glj, Hasegawa:2019jsa} and can be computed from the definition of heating efficiency in \eqref{eq:heating_eff} and heating temperature in \eqref{eq:heating_temp}. A maximum scale is set by the requirement of sub-Planckian non-minimal-coupling contributions, i.e. guaranteeing the subdominance of the spectator field and the consistency of the analysis. In Figure \ref{fig:GW_detectors_hkin}, we show the full GW signal for different scales of kination. In this case, for $\nu=10$ and $\lambda=10^{-10}$ the kination scale has to be in the range ${10^5 \textrm{ GeV}\lesssim H_{\rm kin} \lesssim 10^{11} \textrm{ GeV}}$.

The low-frequency tail in Figure \ref{fig:GW_detectors_strain} shows the unavoidable contribution to the stress-energy tensor given by perturbations generated during inflation from quantum fluctuations. Being this source uncorrelated to the HIPT one, we can simply compute the inflationary GW power spectrum and add it to the one generated by the spectator field's fluctuations. It is well-known \cite{Allahverdi:2020bys, Chen:2024roo, Opferkuch:2019zbd} that a stiff expansion phase following inflation leads to an amplification of the GW modes over the cosmological background and a blue-tilting of the spectrum. This is described in terms of two different scalings. For modes reentering during the phase of radiation-domination, i.e. $f < f_{\rm ht}$ the power spectrum nowadays is described by
\begin{equation}
    \Omega_{\text{GW, 0}}^{\rm inf}(f) \simeq 10^{-16}\left(\frac{H_{\rm kin}}{ H_{\rm max}}\right)^2\left(\frac{f}{f_{\rm pivot}}\right)^{n_t} \; ,
\end{equation}
while for modes reentering during the stiff phase of kination, i.e. $f > f_{\rm ht}$, we rather have
\begin{equation} 
    \Omega_{\text{GW, 0}}^{\rm inf}(f) \simeq 10^{-16}\left(\frac{H_{\rm kin}}{ H_{\rm max}}\right)^2\left(\frac{f}{f_{\rm pivot}}\right)^{n_t} \left({\frac{f}{f_{\rm ht}}}\right) \; ,
\end{equation}
where $H_{\rm max}$ is the highest inflationary scale measured at the pivot scale $k_{\rm pivot}=0.002 \text{ Mpc}^{-1}$ \cite{Planck:2018jri}, $f_{\rm pivot}$ is the corresponding frequency of the pivot scale and $n_t$ is the spectral tilt of inflationary perturbations. The range of frequencies that experience such amplification is limited to $f>f_{\rm ht}$, where $f_{\rm ht}$ is the frequency of the mode exiting the horizon at the end of heating. The typical amplitude of such contribution is negligible as compared to the HIPT signal, unless the anisotropies generated during the tachyonic phase do not achieve a large amplitude and the kination phase is long-lasting. This is the case that can be observed in Figure \ref{fig:GW_detectors_strain} for low kination scales, i.e. small heating efficiencies, where we have converted amplitudes into stochastic strain via the standard relation \cite{Aggarwal:2020olq}
\begin{equation}
    \Omega_{\rm GW}=\frac{4\pi^2}{3H_0^2}f^2h_c^2(f) \;.
\end{equation} 
As Figures \ref{fig:GW_detectors_lambda}, \ref{fig:GW_detectors_hkin} and \ref{fig:GW_detectors_strain} show, the GW signal is peaked at typical frequency scales far from the reach of most ground-based and space-based detectors. This is a core feature of peaked violent processes taking place at high energy scales. Moreover, the peculiar scaling features of our scenario limit the possibility of reducing the typical frequency below the MHz threshold. At the present time, high-frequency detector designs \cite{Gatti:2024mde, Aggarwal:2020olq} have the potential to reach the MHz-GHz range, but only with maximum sensitivities far above the bound imposed by $\Delta N_{\rm eff}$ at BBN. 

\section{Conclusion and outlook} \label{sec:conclusion}

The present work has filled in a gap in the study of primordial fields undergoing a Hubble-induced phase transition. The building blocks of the model are minimal and require only the presence of non-minimally-coupled spectator fields in the early Universe as well as a stiff expansion phase. Any model satisfying these two assumptions experiences such phase transition which, for values of the non-minimal coupling parameter $\xi$ of the order $10^1-10^3$, can lead to the successful heating of the Universe with an associated gravitational-wave signal. By performing a large number of $3+1$--dimensional classical lattice simulations, we have thoroughly investigated the stochastic background of gravitational waves produced by such phase transition. We have summarised our findings in a set of parametric fitting formulas that conveniently encode the main characteristic quantities: the integrated power spectrum, its peak amplitude and the corresponding frequency. We have also provided the first characterisation of the spectral shape for a crossover phase transition using a broken power-law fit.

The advantages of such parametric formulas are clear: key quantities can be estimated without the need for more time-consuming simulations. Moreover, such results can be taken as starting points for extending the analysis to neighbouring scenarios involving, for example, a different stiff equation of state, additional field content and interactions \cite{Repond:2016sol,Fan:2021otj}, additional non-standard expansion phases etc. In particular, the early-Universe evolution of the Standard Model Higgs field considered in \cite{Laverda:2024qjt} represents a natural application of our results, offering the chance to correlate the gravitational-wave spectrum to the running of the Higgs self-coupling at high energies. This new connection can lead to an interesting interplay between cosmological and particle-physics measurements, in particular the top quark mass. We plan on further investigating this scenario in a follow-up work.  

\acknowledgments
G.~L. wishes to thank Matteo Piani for the interesting discussions during the preparation of this work. The numerical lattice simulations have been performed with the support of the Infraestrutura Nacional de Computa\c c\~ao Distribu\'ida (INCD) funded by the Funda\c c\~ao para a Ci\^encia e a Tecnologia (FCT) and FEDER under the project 01/SAICT/2016 nº 022153. G.~L. acknowledges support from a fellowship provided by ``la Caixa” Foundation (ID 100010434) with fellowship code LCF/BQ/DI21/11860024, as well as the support of FCT through the grant with Ref.~2024.05847.BD. G.~L. thanks also FCT for the financial support to the Center for Astrophysics and Gravitation-CENTRA, Instituto Superior T\'ecnico,  Universidade de Lisboa, through the Project No.~UIDB/00099/2020. D.~B. acknowledges support from Project PID2021-122938NB-I00 funded by the Spanish “Ministerio de Ciencia e Innovación” and FEDER “A way of making Europe” and Project SA097P24 funded by Junta de Castilla y León.  D.~B. and J.~R.~acknowledge support from Project PID2022-139841NB-I00 funded by the Spanish “Ministerio de Ciencia e Innovación”. J.~R. is supported by a Ram\'on y Cajal contract of the Spanish Ministry of Science and Innovation with Ref.~RYC2020-028870-I. A.L-E acknowledges support from Eusko Jaurlaritza (IT1628-22) and by the PID2021-123703NB-C21 grant funded by MCIN/AEI/10.13039/501100011033/ and by ERDF; “A way of making Europe”.

\appendix

\section{Derivation of the source term for gravitational waves} \label{app:GW_source}

\begin{figure}[tb]
\centering
\includegraphics[width=0.8\textwidth]{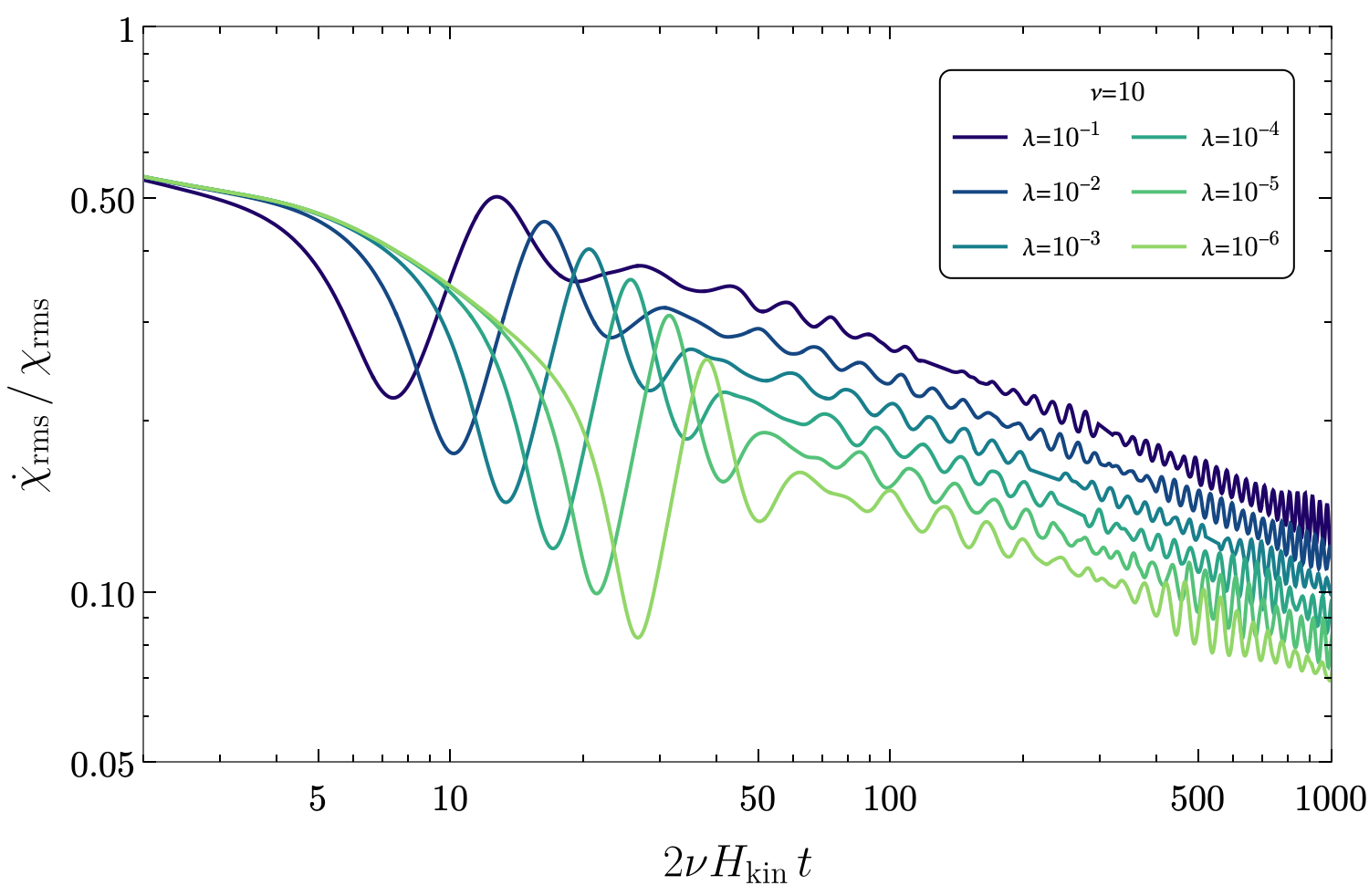}
\hspace{7mm}
\caption{Time-evolution of the ratio $\Dot\chi_{\rm rms}/\chi_{\rm rms}$, where a prime indicates a derivative with respect to $z$. Different colours indicate the numerical output of different lattice simulations with constant $\nu=10$. } \label{fig:chi_ratio}
\end{figure}

In this Appendix, we present a detailed derivation of the GW equations \eqref{eq:TT_stress_energy_tensor}. To this end, we employ the spectator field approximation, specifically assuming that the dominant inflaton field remains strictly homogeneous. Under this assumption, the spatial components of the perturbed Einstein equations take the form
\begin{equation}\label{pertG}
\left(1-\frac{\xi \chi^2}{M_P^2}\right) \delta G_{ij}= \frac{1}{M_P^2} \delta\tilde \Pi_{ij}\,,
\end{equation}
with
\begin{equation}
\delta\tilde \Pi_{ij} \equiv \tilde  \Pi_{ij} - \frac{1}{3} g_{ij} \tilde \Pi  = \partial_i \chi \partial_j \chi- \frac{1}{3} g_{ij}  \partial^k \chi \partial_k\chi  - \xi \left(  \nabla_i \nabla_j-  \frac{1}{3} g_{ij}  \nabla^k \nabla_k \right) \chi^2 
\end{equation}
an anisotropic stress tensor quantifying the deviation of the spatial--spatial components of the ``reduced" energy-momentum tensor,
\begin{equation}
\tilde \Pi_{\mu\nu}=\partial_\mu \chi \partial_\nu \chi - g_{\mu\nu} \left( \frac{1}{2} \partial^\lambda \chi \partial_\lambda \chi + \frac{\lambda}{4}\chi^4 \right) + \xi \left( g_{\mu\nu} \Box - \nabla_\mu \nabla_\nu \right) \chi^2\,,
\end{equation}
from a perfect fluid. Particularising this expression for a perturbed FLRW Universe,  
\begin{equation}
g_{ij}=a^2\gamma_{ij}=a^2 (\delta_{ij}+h_{ij})
\,, \quad  \nabla_i \nabla_j \chi = \partial_i \partial_j \chi - \left({\Gamma}{}^\sigma{}_{ij}+\delta_h\Gamma^\sigma{}_{ij}\right)\partial_\sigma \chi\,, \quad {\Gamma}{}^0{}_{ij}=a^2 H \delta_{ij}\,, \nonumber 
\end{equation}
we get
\begin{multline} \label{Piexpanded}
\delta\tilde \Pi_{ij} = \partial_i \chi \partial_j \chi- \frac{1}{3} a^2 \gamma_{ij}  \partial^k \chi \partial_k\chi  \\ - \xi \left(  \partial_i \partial_j - {\Gamma}{}^0{}_{ij}\partial_0-\delta_h\Gamma^0{}_{ij} \partial_0- \delta_h\Gamma^k{}_{ij} \partial_k -  \frac{1}{3} a^2 \gamma_{ij} \nabla^k \nabla_k \right) \chi^2  \,. 
\end{multline} 
The evolution equation for the physical transverse-traceless gravitational degrees of freedom $h_{ij}^{TT}$ actually propagating and carrying energy away from the source is derived by first transforming \eqref{pertG} into Fourier space and then applying spatial projectors \( P_{ij} = \delta_{ij} - \hat{k}_i \hat{k}_j \), with \( \hat{k}_i = k_i / k \) the unit vector along the wave vector direction. This procedure removes all pure trace of components in \eqref{Piexpanded}, and in particular those associated with the $\delta_{ij}$ part in $\gamma_{ij}$ and the background Christoffel symbols ${\Gamma}{}^0{}_{ij}=a^2 H \delta_{ij}$. Further neglecting metric backreaction terms of the form $h_{ij}\partial^k\chi \partial_k\chi$, $\delta_h\Gamma^\sigma{}_{ij} \partial_\sigma\chi^2$ and $h_{ij}\partial^k\partial_k\chi^2$ in consistency with the spectator field approximations $\xi\chi^2/M_P^2\ll1$ and $(\partial\chi)^2 \ll H^2_{\rm kin}M_P^2$, and transforming the result back into position space, the TT part of perturbed Einstein equations \eqref{pertG} takes the approximate form \eqref{eq:omega_GW_definition}, namely
\begin{equation}
\ddot h_{ij}^{TT}+ 3H\dot h_{ij}^{TT} - \frac{1}{a^2}\nabla^2 h_{ij}^{TT}\simeq \frac{2}{a^2M_P^2}  {\Pi}^{TT}_{ij}\,, 
\end{equation}
with ${\Pi}^{TT}_{ij}$ the TT part of the effective stress tensor $
\Pi_{ij} \equiv \partial_i\chi\partial_j\chi  - \xi\partial_i\partial_j\chi^2$. Note in particular that the contribution arising from the term proportional to $\Dot\chi/\chi$ remains much smaller than the Hubble scale as long as $\xi \chi^2/M_P^2\ll1$. This is illustrated in Figure \ref{fig:chi_ratio}, which displays the lattice-averaged rms quantities.

\bibliographystyle{JHEP.bst}
\bibliography{Kin-heat}
\end{document}